\documentstyle[11pt]{article}

\relax
\catcode`\@=11

\newcommand{\e}{{\rm e}}                               
\newcommand{\ii}{{\rm i}}                               
\newcommand{\dd}{{\rm d}}                                
\newcommand{\half}{{1 \over 2}}                           

\newcommand{\Lie}{{\cal L}}
\newcommand{\MM}{{\cal M}}
\newcommand{\RR}{{\cal R}}

\newcommand{\DD}{{\cal D}}
\newcommand{\QQ}{{\cal Q}}
\newcommand{\FF}{{\cal F}}
\newcommand{\A}{{\cal A}}
\newcommand{\B}{{\cal B}}
\newcommand{\CC}{{\cal C}}

\newcommand{\HH}{{\cal H}}

\newcommand{\doo}{{\partial}}
\newcommand{\sgn}{{\rm sign\  } }

\newcommand{\Ch}{{\rm Ch   }  }

\newcommand{\Det}{{\rm Det\ }}
\newcommand{\Pf}{{\rm Pf \  }}
\newcommand{\Str}{{\rm Str \  }}

\newcommand{\ra}{\rightarrow}

\newcommand{\be}{\begin{equation}}
\newcommand{\ee}{\end{equation}}

\newcommand{\ba}{\begin{eqnarray}}
\newcommand{\ea}{\end{eqnarray}}

\begin{document}

\begin{titlepage}
\begin{flushright}
UU-ITP 14-1996 \\ hep-th/9606083
\end{flushright}

\vskip 0.5truecm

\begin{center}
{ \large \bf
Antibrackets, Supersymmetric $\sigma$-Model and Localization
\\  }
\end{center}

\vskip 1.0cm

\begin{center}
{\bf Mauri Miettinen$^*$ \\
\vskip 0.4cm
{\it Department of Theoretical Physics, Uppsala University
\\ P.O. Box 803, S-75108, Uppsala, Sweden \\}
\vskip 0.4cm}
\end{center}

\vskip 1.0cm

\rm
\noindent
We consider supersymmetrization of Hamiltonian dynamics via
antibrackets for systems whose Hamiltonian generates an isometry of
the phase space. We find that the models are closely related to the
supersymmetric non-linear $\sigma$-model. We interpret the
corres\-ponding path integrals in terms of super loop space equivariant
cohomology. It turns out that they can be evaluated exactly using
localizations techniques.

\vfill

\begin{flushleft}
\rule{5.1 in}{.007 in}\\
$^{*}${\small E-mail: $~~$ {\small \bf
mauri@rhea.teorfys.uu.se} $~~~~$
}
\end{flushleft}

\end{titlepage}

Localization theorems (for a review see eg. \cite {Blau}) state that
certain infinite or finite dimensional integrals can be reduced to
finite dimensional integrals or discrete sums, provided certain
geometrical conditions are satisfied. This happens for example in
topological theories where localization is achieved using Mathai-Quillen
formalism \cite{MQ}. On the other hand, the Duistermaat-Heckman theorem
\cite{dh} can be applied to some Hamiltonian systems if the Hamiltonian
generates isometries on the phase space.  In this letter we shall
consider supersymmetrized Hamiltonian mechanics using antibrackets and
path integral localization techniques. The antibracket was introduced
by Batalin and Fradkin \cite{Batalin} for the Lagrangian quantization
of field theories with constraints. In this letter we use it to define
odd symplectic structures on supermanifolds, which allows us to
consider Hamiltonian systems. However, since the corresponding
Hamiltonian is odd, it is necessary to construct an even symplectic
structure and an even Hamiltonian to quantize the systems with path
integrals. This turns out to be possible for Hamiltonians which
generate isometries \cite{Nersessian}. This condition enables the
application of localization methods \cite{us} based on the equivariant
cohomology \cite{BGV} to evaluate the corresponding path integrals
exactly. These results can be interpreted as infinite dimensional
generalizations of the Duistermaat-Heckman theorem \cite{dh}.

This letter is organized as follows. First we consider formulation of
Hamiltonian dynamics on a supermanifold using both odd and even
symplectic structures. With the even structure we are able to consider
the corresponding quantum mechanical partition functions. It turns out
that the models are closely related to the supersymmetric non-linear
$\sigma$-model. We interpret the pertinent path integrals using
equivariant cohomology and evaluate them exactly by localization. The
results turn out to be equivariant generalizations of familiar
topological invariants.

We start by considering a Hamiltonian system $(M,\omega, H)$ where $M$
is a phase space with a symplectic 2-form $\omega = \half
\omega_{ab} \dd x^a \wedge \dd x^b$ (in local coordinates)
and $H$ is the Hamiltonian.  The symplectic structure determines the
Poisson bracket
\be\label{EvenPoisson}
\{A, B\} = \omega(\chi_A, \chi_B) = \doo_a A \omega^{ab} \doo_b B   \;.
\ee
Hamilton's equations of motion are given by $ \dot{x}^a = \{x^a , H \}
= \chi^a_H$. Here $\chi_H$ denotes the Hamiltonian vector field
corresponding to $H$  determined by the equation
\be
\dd H + \iota_H \omega = 0 \,
\ee
where $\iota_H$ denotes the contraction along $\chi_H$.

We can also consider Hamiltonian dynamics on the supermanifold $\MM$
associated to the cotangent bundle $T^* M$. The coordinates on $\MM$
are denoted by $z^A =(x^a, \theta^a)$. We define an odd symplectic
structure on $\MM$ by introducing a non-degenerate odd symplectic
2-form \be\label{OddSymplectic} \Omega^1 = \dd z^A \Omega^1_{AB} \dd
z^B \ee This 2-form determines the antibracket (odd Poisson bracket)
whose grading and antisymmetry properties are opposite to those of the
ordinary graded Poisson bracket: \ba\label{Grading} \epsilon(\A, \B )
&=& \epsilon (\A) + \epsilon(\B) +1 \;, \cr ( \A, \B) & = & -
(-1)^{(\epsilon_A + 1) (\epsilon_B +1) } (\B, \A) \;,\cr (\A, \B \CC)
& = & (\A, \B) \CC + (-1)^{\epsilon_B (\epsilon_A+1) } \B (\A, \CC)
\cr 0 & = & (-1)^{(\epsilon_A +1)(\epsilon_C +1)} (\A, (\B,\CC)) +
{\rm cyclic\ perm} \;.  \ea The 2-form $\Omega^1$ can be written in
local coordinates $z^A = (x^a, \theta^a) $ as
\be\label{LocalSymplectic} \Omega^1 = \omega_{ab} \dd x^a \wedge \dd
\theta^b + {\doo \omega_{ab} \over \doo x^c} \theta^c \dd \theta^a
\wedge \dd \theta^b \ee and the antibracket becomes
\be\label{Antibracket} (\A,\B ) = \omega^{ab} \left({ \doo^r \A \over
\doo x^a } {\doo^l \B \over \doo \theta^b} - { \doo^r \B \over \doo
x^a } {\doo^l \A \over \doo \theta^b } \right) + { \doo \omega^{ab}
\over \doo x^c } \theta^c {\doo^r \A \over \doo \theta^a } {\doo^l \B
\over \doo \theta^b } \ee where where $\A(x,\theta)$ and $\B(x,
\theta)$ etc. are superfunctions on $\MM$.  The superscripts $r$ and
$l$ denote right and left derivatives, respectively.  In particular,
we have the basic antibrackets \ba\label{BasicBrackets} (x^a, x^b) &=&
0 \;, \cr (x^a , \theta^b ) & = & \omega^{ab} \;, \cr (\theta^a,
\theta^b ) & = & { \doo \omega^{ab} \over \doo x^c } \theta^c \;.  \ea

We define a dynamical system on $\MM$ by mapping the original
Hamiltonian to an odd Hamiltonian with the function $\FF = \half
\omega_{ab} \theta^a \theta^b$
\be\label{OddHamiltonian}
\QQ_H = (H, \FF ) = { \doo H \over \doo x^a } \theta^a \;.
\ee
The antibracket coincides with the original bracket
\be\label{OddIsEven}
(\A , \QQ_H ) = \{\A , H \}  \;.
\ee
 The corresponding equations of motion are
\ba\label{EQM}
\dot{x}^a& =& ( x^a , \QQ_H ) = \chi^a_H  \;, \cr
\dot{\theta^a}& =& ( \theta^a , \QQ_H ) = \doo_b \chi^a_H \theta^b \;.
\ea
Using the antibracket we have thus found a supersymmetric
generalization of the ordinary Hamiltonian dynamics.

{}From now on we assume that the original symplectic manifold $M$ admits
a Riemannian metric $g$ for which $\chi_H$ is a Killing vector:
\be\label{Lie}
\Lie_H g = 0
\ee
or in a component form
\be\label{ComponentForm}
\chi_H^c \doo_c  g_{ab} + \doo_a \chi^c_H g_{bc} + \doo_b \chi_H^c
g_{ac} = 0 \;.
\ee
Then the function $\tilde{\QQ}_H = \half g_{ab} \chi^a_H \theta^b$ is
an integral of motion for the antibracket
\be\label{IOM}
(\QQ_H, \tilde{\QQ}_H ) =0 \;.
\ee
$\tilde{\QQ}_H$ also yields a bi-Hamiltonian structure on $M$ with the
second symplectic structure $\tilde{\omega}$ and Hamiltonian $H_2 = K$
\ba\label{Bi-Hamiltonian}
( \FF, \tilde{\QQ}_H ) &=& \half \tilde{\omega}_{ab} \theta^a
\theta^b = \half \left[ \doo_a (g_{bc} \chi_H^c ) - \doo_b (g_{bc}
\chi_H^c ) \right] \theta^a \theta^b \;,\cr
( H, \tilde{\QQ}_H ) &=& H_2 = \half g_{ab} \chi^a_H \chi^b_H
\ea
Assuming that $\tilde{\omega}$ is non-degenerate we see that the
equations of motions coincide
\be\label{Integrability}
{\doo K \over \doo x^a} = \tilde{\omega}_{ab} \omega^{bc} {\doo H
\over \doo x^c}
\ee
This means that the system is classically integrable.

We proceed to reformulate the dynamics on $\MM$ using an even Poisson
bracket. We want to do this since we need an even Hamiltonian and
symplectic structure for path integral quantization. An even
symplectic structure on $\MM$ is given by the following
supersymplectic 2-form \cite{Nersessian}
 \be\label{EvenForm}
\Omega_{\alpha} = \half \left( \omega_{(\alpha)ab} + R_{abcd} \theta^c
\theta^d \right) \dd x^a \wedge \dd x^b + \half g_{ab} D \theta^a
\wedge D \theta^b \;.
\ee 
Here $R$ is the Riemannian curvature of $M$ and
$D \theta^a = \dd \theta^a + \Gamma^a_{bc} \theta^b \dd x^c$ the
covariant derivative on $M$ with the metric connection
\be\label{Christoffel} \Gamma^a_{bc} = \half g^{ae} \left( \doo_b
g_{ce} + \doo_c g_{be} - \doo_e g_{bc} \right) \;.  \ee The subscript
$\alpha = 0,2$ refers to the symplectic structures $(H_0 = H ,\omega_0
= \omega, M)$ and $(H_2, \omega_2 = \tilde{\omega}, M)$. The
corresponding symplectic 1-forms are \ba\label{OneForms} \Theta_0 &=&
\vartheta_a \dd x^a + g_{ab} \theta^a D\theta^b \cr \Theta_2 &=&
g_{ab}\chi_H^b \dd x^a + g_{ab} \theta^a D\theta^b \ea The 2-form
$\Omega_{\alpha}$ determines the even Poisson bracket on $\MM$
\be\label{EvenPoisson2} [ \A, \B ]_{\alpha} = \nabla_a \A \Vert
\omega_{ab} + R_{abcd} \theta^c \theta^d \Vert^{-1} \nabla_b \B +
g^{ab} {\doo^r \A \over \doo \theta^a} {\doo^l \B \over \doo \theta^b}
;, \ee 
where 
\be\label{Covariant} \nabla_a = \doo_a - \Gamma^b_{ac}
\theta^c {\doo \over \doo \theta^b} \;.  \ee The equations of motion
for odd and even Poisson brackets on $\cal{M}$ coincide:
\be\label{Coincide} [ z^A, {\cal \HH_{\alpha} ]_{\alpha} } = ( z^A ,
\QQ_H \ ) \ee where $\HH_{\alpha} = H_{\alpha} + \tilde{\omega} $. The
odd and even Poisson brackets provide a bi-Hamiltonian structure also
on $\MM$ and therefore the system is integrable also on $\MM$.

Having found an even symplectic structure on $\MM$ we proceed to
quantize these systems by path integrals. We are interested in
evaluating the partition functions of the form \be\label{Z} Z = \Str
\exp[-\ii TH] = \int \DD \mu \exp[\ii S] \ee where $H$ and $S$ are the
pertinent Hamiltonians and classical actions, respectively and $\DD
\mu$ denotes integration over the (super) loop space $L M$ which
consists of all $T$-periodic loops for both commuting and
anticommuting variables. The integration measure is the Liouville
measure determined by the symplectic structure (\ref{EvenForm}).  The
system $(M, H, \omega)$ has the classical action \be\label{Action} S =
\int_0^T \dd t [ \vartheta_a \dot{x}^a -H ] \ee and the corresponding
partition function becomes \be\label{FPI} Z = \int \DD x \Pf \Vert
\omega_{ab} \Vert \exp[\ii S ] = \int \DD x \DD \eta \exp[\ii
\tilde{S} ] \ee where we have introduced anticommuting variables
$\eta$ write the Liouville measure factor $\Pf \Vert \omega_{ab}
\Vert$ as path integral. Now have the following action
\be\label{SUSYAction} \tilde{S}_0 = \int_0^T \dd t \left[ \vartheta_a
\dot{x}^a -H + \half \omega_{ab}\eta^a \eta^b \right] \;.  \ee

The systems $(\MM , H_{\alpha} , \Omega_{\alpha} )$ have the classical
actions \ba\label{SUSYClassical} S_{0} &= & \int_0^T \dd t [
\vartheta_a \dot{x}^a - H + \half \theta^a g_{ab} {D \theta^b \over
\dd t} + \half \tilde{\omega}_{ab} \theta^a \theta^b ] \cr S_2 & = &
\int_0^T \dd t [ g_{ab} \chi_H^b \dot{x}^a -\half g_{ab} \chi_H^a
\chi_H^b + \half \theta^a g_{ab} { D \theta^b \over \dd t } +
\tilde{\omega}_{ab} \theta^a \theta^b ] \ea where \be\label{CurveCov}
{D \theta^b \over \dd t } = {\dd \theta^b \over \dd t} + \dot{x}^d
\Gamma^b_{dc} \theta^c \ee is the covariant derivative along the curve
$x(t)$.  The corresponding path integral becomes, using the superspace
Liouville measure determined by $\Omega_{\alpha}$ \ba \label{SUSYFPI}
Z_{\alpha} = \int \DD x \DD \theta {{\Pf \Vert \omega_{(\alpha)ab} +
R_{abcd} \theta^c \theta^d \Vert } \over \sqrt{\Det \Vert g_{ab} \Vert
}} \exp [ \ii {S}_{\alpha} ] = \int \DD x \DD \theta \DD F \DD \eta
\exp [ \ii \tilde{S}_{\alpha}] \ea where $\tilde{S}_{\alpha}$ denotes
now the quantum actions \ba\label{QuantumActions} \tilde{S}_0 & =&
\int_0^T \dd t [ \vartheta_a \dot{x}^a - H + \half \theta^a g_{ab} {D
\theta^b \over \dd t} + \half \tilde{\omega}_{ab} \theta^a \theta^b +
\half \omega_{ab} \eta^a \eta^b \cr &+ & \half R_{abcd} \eta^a \eta^b
\theta^c \theta^d + \half g_{ab} F^a F^b ] \cr \tilde{S}_2 & = &
\int_0^T \dd t [ g_{ab} \chi_H^b \dot{x}^a - \half g_{ab} \chi^a_H
\chi^b_H + \half \theta^a g_{ab} {D \theta^b \over \dd t} + \half
\tilde{\omega}_{ab} \theta^a \theta^b + \half \tilde{\omega}_{ab}
\eta^a \eta^b \cr &+& \half R_{abcd} \eta^a \eta^b \theta^c \theta^d +
\half g_{ab} F^a F^b ] \ea Here we have again introduced anticommuting
variables $\eta^a$ and commuting ones $F_a$ to exponentiate the
determinants. In particular, if $H=0$ this almost reduces to the $N=1$
nonlinear supersymmetric $\sigma$-model, up to kinetic term $\half
g_{ab} \dot{x}^a \dot{x}^b which is missing.$ We shall discuss the
relation of this action to the non-linear $\sigma$-model later. In
addition we have coupling to symplectic potentials $\vartheta_a$ and
$g_{ab} \chi_H^b$ which can be interpreted as external gauge fields.

We shall now interpret the path integrals using super loop space
equivariant cohomology. In the following we concentrate only on the
path integrals for $(M, \omega, H)$ and $(\MM, \Omega_0, \HH)$. The
reasoning is similar for the action $\tilde{S}_2$ with the
replacements $H \ra K$ and $\omega \ra \tilde{\omega} $.  To do this
we introduce exterior derivatives in $L \MM$.  Half of the variables
are interpreted as coordinates and the other half as
1-forms. Parameter integrations will not be explicitly written. The
first derivative is the equivariant exterior derivative in the loop
space relevant to the path integral for (\ref{SUSYAction}) whose
bosonic part determines a loop space vector field $\chi_S =
\dot{x} -\chi_H$ whose zeroes define the Hamilton's equations of
motion. We identify the anticommuting variables $\eta$ as 1-forms and
define the basis for contractions
\ba\label{Basis}
\eta^a &\sim& \dd x^a \;, \cr
\iota_a \eta^b & =& \delta_a^b \;.
\ea
We define the loop space equivariant exterior derivative
\be\label{EqDer}
\dd_S = \dd + \iota_S = \eta^a {\delta \over \delta x^a } +
\chi_S^a \iota_a = \eta^a  {\delta \over \delta x^a }
+ (\dot{x}^a - \chi_H^a) \iota_a
\ee
which squares to the loop space Lie derivative $\Lie_S \sim \dd / \dd
t - \Lie_H$. This is effectively nilpotent on periodic loops and on
the invariant subcomplex of equivariant differential forms. We may use
the freedom to make canonical transformations for the symplectic
potential $\vartheta \ra {\vartheta} + \dd
\psi$ such that  $ \Lie_H \vartheta = 0$.  Using this
we can identify $H= \iota_H \vartheta$ which allows us to write the
entire action in a form of a cohomological theory
\be\label{Coh}
\tilde{S} = \dd_S \vartheta \;.
\ee

The other exterior derivative is a proper super loop space exterior
derivative (truly nilpotent) obtained by considering $x, \eta$ as
coordinates and $\theta, F$ their differentials
\be\label{SuperDer}
\delta =\theta^a {\delta \over \delta x^a } +
F^a {\delta \over \delta \eta^a } \;.
\ee
Now the actions (\ref{QuantumActions}) are obtained by $\tilde{S}_0 =
\tilde{S} + \delta \Theta$ with
 \be\label{GaugeFermion}
\Theta = \half g_{ab} (\dot{x}^a - \chi^a ) \theta^a  + \half \Gamma_{bc}^a
\eta^b \eta^c \theta_a + \half g_{ab} \tilde{F}^a \eta^b
\ee
where $\tilde{F}^a = F^a -\Gamma^a_{bc} \eta^b \theta^c$.  On the
classical level this is consistent with the Poisson bracket relations
on $M$ and $\cal{M}$ since the addition of an (locally) exact piece
should not change the equations of motion obtained from the action
principle. However, the path integrals will be different since we have
a different number of degrees of freedom.  

Now we shall first evaluate the path integrals exactly using
localization techniques \cite{us} based on equivariant cohomology in
super loop space. The evaluation relies on the Lie-derivative condition
for the metric. We first concentrate on the path integral
(\ref{SUSYAction}). The action $\tilde{S}$ is $\dd_S$-closed and
therefore the path integral remains intact under $\tilde{S} \ra
\tilde{S} + \lambda \dd_S \Psi$ whenever the 1-form (gauge fermion)
$\Psi$ satisfies $\dd_S^2 \Psi = \Lie_S \Psi = 0 $.  The limit
$\lambda \ra 0$ gives the original path integral and $\lambda \ra
\infty$ gives a localization to certain configurations, depending on $\Psi$.

To construct a gauge fermion $\Psi$ we also need a metric in
$M$. Under the assumption that $\chi_H$ is a Killing vector for
$g_{ab}$ the following gauge fermions satisfy (\ref{Lie}), as we have
assumed. $\Psi = \half g_{ab} \dot{x}^a \eta^b$ reduces the path
integral to an ordinary integral over $M$, producing the result
\be\label{EQIndex}
Z = \int \dd x \dd \eta \exp[ - \ii T(H- \omega)] \sqrt{\Det
\left[ {
\RR /2 \over  \sin (T \RR /2 ) } \right] }= \int_M \Ch[- \ii T(H-\omega) ]
\hat{A} [T \RR ]
\ee
where $\RR = \Vert R^a_b + \tilde{\Omega}^a_b \Vert$ is the
equivariant curvature of $M$. The symbols $\Ch$ and $\hat{A}$ denote
the (equivariant) Chern class and the Dirac genus. 
To obtain this result we have separated $x, \eta$ to constant and
non-constant modes and scaled the latter ones
\ba\label{Scaling}
x &=& x_0 + { 1 \over \sqrt{\lambda}} x_t \cr
\eta &=& \eta_0 + {1 \over \sqrt{\lambda}} \eta_t
\ea
and integrated over non-constant modes. The Jacobian for this
transformation is trivial. The expression (\ref{EQIndex}) is an
equivariant generalization for the Atiyah-Singer index for a Dirac
operator on a Riemannian manifold, reducing to the standard index when
$H=0$.

We shall now consider the path integral for the system $(\MM,
\Omega_{\alpha}, H_{\alpha} )$ \cite{sigma}. We interpret $x^a$ and
$\theta_a$ as coordinates in the super loop space and $\eta^a$ and
$F_a$ as their differentials and introduce corresponding contractions
\ba\label{SUSYContractions} \eta^a &\sim& \dd x^a \:, \cr F_a &\sim &
\dd \theta_a \;, \cr \iota_a \eta^b &=& \delta_a^b \cr \pi^a F_b & = &
\delta_b^a \;.  \ea We define the super loop space equivariant
exterior derivative \ba\label{SuperEQ} Q = \eta^a {\delta \over \delta
x^a } + F_aa {\delta \over \delta \theta_a } + (\dot{x}^a - \chi^a_H )
\iota_a + (\dot{\theta}_b + \doo_b \chi^a_H \theta_a ) \pi^b \; \ea
with $Q^2 =\Lie_S = \dd / \dd t - \Lie_H$.  We introduce a canonical
conjugation $Q \ra \e^{-\Phi} Q_S \e^{\Phi} = \tilde{Q}$ which does
not change its cohomology. Choosing the functional $\Phi =
\Gamma^a_{bc} \pi^b\eta^c\theta_a$ we get \ba\label{Conjugation}
\tilde{Q} &=& \eta^a {\delta \over \delta x^a } + F_a {\delta \over
\delta \theta_a } + (\dot{x}^a - \chi^a_H ) \iota_a + (\dot{\theta}_a
+ \doo_a \chi^b_H \theta_b ) \pi^a \cr &+& \left( \Gamma^a_{bc} F_a
\eta^b - \half R^a_{bcd} \eta^c \eta^d \theta_a - (\dot{x}^b -
\chi^b_H) \Gamma^a_{bc} \theta_a + (\delta^a_b \doo_t + \doo_a
\chi_H^b ) \theta_c \right) \pi^c \;.  \ea The pertinent action
(\ref{QuantumActions}) can be obtained from the 2-dimensional $N=1$
supersymmetric $\sigma$-model by partial localization to a
1-dimensional model \cite{sigma}. In this procedure the kinetic term
for the bosons (in light-cone coordinates) $ g_{ab} \doo_+ \phi^a
\doo_- \phi^b$ drops out.  The action $\tilde{S}_0 $ is
$\tilde{Q}$-closed, $\tilde{Q} \tilde{S}_0 = 0$, and therefore the
path integral $Z_0$ remains intact if we add a term $\tilde{Q} \Psi $
to the action. We shall consider the following gauge fermions.
$\Psi_1 = g_{ab} F^a \theta^b + {\lambda \over 2} g_{ab} (\dot{x}^a -
\chi^a_H ) \eta^b$ localizes the path integral in the limit $\lambda
\ra \infty$ to the $T$-periodic classical trajectories for the
original action (\ref{Action}) \be\label{Poincare-Hopf} Z_0 =
\sum_{\delta S = 0} \sgn \left[\Det \Vert {\delta^2 S \over \delta x^a
\delta x^b } \Vert \right] \exp[\ii S ] \ee which can be interpreted
as a loop space generalization of the Poinc\'are-Hopf theorem. By
selecting $\Psi_2 = g_{ab} F^a \theta^b+ {\lambda \over 2 } g_{ab}
\dot{x}^a \eta^b$ we find that the path integral localizes to an
ordinary integral over $M$ \ba\label{Gauss-Bonnet} Z_0 &=& \int \dd x
\dd \eta \exp[- \ii T(H- \half \omega_{ab} \eta^a \eta^b ) ] \Pf
\left[\half (\tilde {\omega}^a_b + R^a_{bcd} \eta^c \eta^d ) \right]
\cr & =& \int_{M} \Ch[-\ii T(H- \omega)] {\rm Eul} \left[ \RR \right]
\ea which is an equivariant generalization of the Gauss-Bonnet theorem
for the Euler charac\-teristic of $M$. Indeed, setting, $H= \vartheta
=0$ we obtain the standard formula \be\label{Euler} \chi(M) = \int_M
{\rm Eul} (R) = \sum_{\dd H = 0} \sgn [ \det \Vert \doo_a \doo_b H
\Vert ] \;.  \ee

We now discuss some aspects of out results. The localization of the
path integral (\ref{Z}) is related to the loop space generalization of
the Duistermaat-Heckman theorem, to equivariant index and Lefschetz
fixed point theorems. For example, applications to path integral
proofs for index theorems have been considered in \cite{Kaupo}. In
particular, it can be used to evaluate exactly the path integral
related to the quantization of coadjoint orbits of semisimple Lie
groups. In \cite{Mauri} it was shown to exactly yield the correct
character for SU(2) without the Weyl shift by replacing the
$\hat{A}$-genus by Todd-genus.

In summary, we have studied Hamiltonian systems on ordinary symplectic
manifolds and supermanifolds. On supermanifolds we found both odd and
even symplectic structures for Hamiltonians generating isometries. The
existence of  bi-Hamiltonian structures  implied the classical
integrability of the systems. The models on the supermanifold also
turned out to be closely related to the supersymmetric non-linear
$\sigma$-model. It was possible to quantize the systems using the even
symplectic structure and evaluate the path integrals exactly.  The
results were equivariant generalizations of familiar topological
invariants.
\\

{\bf Acknowledgements}

We thank A. Nersessian for initiating this study and useful
discussions and A.J. Niemi for comments on the manuscript.

\end{document}